# Discrete Dislocation Dynamics for Crystal RVEs. Part 1: Periodic Network Kinematics

Yash Pachaury [1], Giacomo Po [2], Anter El-Azab [1]

[1] School of Materials Engineering, Purdue University, West Lafayette, IN

[2] University of Miami, Mechanical and Aerospace Engineering Department, Coral Gables FL

**Abstract**

A novel implementation of the dislocation flux boundary condition in discrete dislocation dynamics is presented. The continuity of the individual dislocation loops in a periodic representative crystal volume (RVE) is enforced across the boundary of the RVE with the help of a dual topological description for representing dislocation line kinematics in two equivalent spaces representing the deforming crystal, the RVE and the unbounded crystal spaces. The former describes the motion of the dislocations in the simulated crystal RVE whereas the latter represents the motion of dislocations in an infinite space containing all replicas of the RVE. A mapping between the two spaces forms the basis of the implementation of flux boundary condition. The implementation details are discussed in the context of statistical homogeneity of bulk crystals undergoing macroscopically homogeneous plastic deformation. Some test cases are presented and discussed to establish the proposed approach. It has been observed that the boundary nodes associated with dislocation segments bear no relevance in the motion of the dislocation segments from the standpoint of bulk crystal plasticity modeling throughout the deformation history of the crystal.

Email Addresses: ypachaur@purdue.edu (Yash Pachaury), aelazab@purdue.edu (Anter El-Azab), gpo@miami.edu (Giacomo Po)

# 1. Introduction

Plastic deformation of crystalline solids is a manifestation of the collective activity of dislocations. Dislocations evolve under the influence of applied loads and their mutual interactions at both short and long ranges. Understanding plastic deformation of crystals based on the fundamentals of dislocation mechanics and physics is a long-standing challenge in materials science. Driven by materials engineering and design needs, theoretical efforts have resulted in the development of a variety of mesoscale models for characterizing the collective behavior of dislocations. While early efforts focused upon a continuum field theory of dislocations (Kröner, 1981), the development of the continuum dislocation dynamics (CDD) method has proven to be extremely challenging, although significant progress has been achieved recently (El-Azab and Po, 2020). On a parallel front, discrete dislocations dynamics (DDD) simulation methodologies have been developed where dislocations are represented as ensembles of lines. The collective efforts of the research community have resulted in mature DDD codes that are capable of emulating the plastic deformation response of metals (Arsenlis et al., 2007; Bulatov et al., 2006; Capolungo and Taupin, 2019; Devincre et al., 2011, 1992; Ghoniem et al., 2000; Kubin et al., 1992; Schwarz, 1998; Verdier et al., 1998; Zbib et al., 2003). Dislocation line discretization options (node-based, lattice based, and parametric), implementation of boundary conditions, dislocation motion, driving force calculation schemes, and implementation of phenomenological rules for short-range interactions and cross slip distinguish different DDD approaches available to the community. Detailed discussions regarding the method of DDD along with its capabilities can be found in (Groh and Zbib, 2009; Kubin, 2013; Marian et al., 2018; Sills et al., 2016).

Accurate formulation and implementation of boundary conditions (BCs) is a very crucial aspect for ensuring that the DDD methodology mimics the mesoscale deformation response of metals. In the last two decades, significant progress has been made for establishing accurate boundary conditions for simulating plasticity in infinite, finite as well as bulk crystals which require infinite, heterogeneous, and periodic BCs, respectively (Van der Giessen and Needleman, 1995; Weygand et al., 2002; Sills et al., 2016). The former case is one in which dislocations are localized in a finite region in space with remote stress boundary conditions. The finite crystal case in one in which one or more dimensions of the crystal end at a physical surface subjected to traction and/or displacement boundary condition. We are interested in the third case, DDD simulation of bulk crystal RVEs, and hence in the formulation and implementation of periodic boundary conditions, henceforth referred to as RVE boundary conditions. An RVE can be defined as the smallest crystal volume whose response is representative of the response of a macroscopic crystal, such that the fields of interest averaged over the RVE at the mesoscale represent the local values in the macroscopic bulk crystal. RVE boundary conditions for the DDD simulation problem must thus enforce specific average fields to the simulated RVE. The technique is similar to what has been proposed by El-Azab (2000), where boundary conditions were derived within the context of CDD for probing the deformation behavior of bulk crystals. The derivations consisted of two necessary and sufficient conditions: the first requires accounting for the stresses due to dislocations which are outside of the simulated crystal volume, and the second requires reinserting the line length lost due to dislocations exiting the boundaries of the crystal RVE. In DDD, the former condition has been satisfied by either using periodic replicas of the crystal RVE in three directions and considering these replicas as images contributing to the RVE stress field after accounting for conditional convergence (Bulatov et al., 2000, Cai et al., 2003) or by solving a well posed elastic boundary value problem (Deng et al., 2008). On the other hand, the flux boundary condition can be taken into consideration by reintroducing the dislocations exiting a surface of the RVE from its periodically opposite surface such that the plastic distortion is preserved for such dislocations (Tang and Kubin, 2001). This paper is focused on the latter specifically on the kinematics of dislocation loops and networks in RVE.

Method of minimum image convention, pioneered by Bulatov et al. (2000, 2006) for DDD, is the most successful and widely used method for enforcing flux continuity across the RVE boundary. In the implementation, the dislocation nodes are present in periodic replicas of the RVE. The minimum image convention works by connecting the two nearest node pairs hence forming a dislocation segment (Bulatov et al., 2006). While the minimum image convention instantaneously enforces flux balance across periodically equivalent boundaries, it requires that all dislocation segments must remain smaller than half of the RVE size (Bulatov et al., 2006, 2000). Significant progress has been made for effectively using the minimum image convention so that spurious annihilations could be alleviated in DDD simulations (Devincre et al. (2011); Madec et al. (2004)). These spurious annihilations are a manifestation of dislocations interacting with their own replicas on periodic reentry. An alternative method for enforcing the flux continuity is through a *kinematic* constraint where the plastic distortion caused by segments exiting from the RVE boundaries need to be tracked and compensated. However, researchers have not had much success in its implementation. The implementation of the kinematic constraint requires tracking the swept area by the dislocations while accounting for various shapes that they can acquire as they exit the RVE boundaries (Tang et al., 2001; Tang and Kubin (2001)). Accounting for short-range interactions such as junction formation and cross-slip increases the complexity of the implementation. Constrained motion of the nodes/segments at the boundary is another challenge in successful implementation of the kinematic constraint.

In this communication, we provide a novel methodology with which the kinematic constraint on dislocations can be instantaneously enforced for satisfying the flux boundary condition. We introduce the RVE concept in DDD, mainly focusing upon the dislocation loop kinematics. The difference between our work and earlier attempts in this regard is that our formulation is based on a dual topological representation of dislocations in two equivalent spaces, namely, the crystal RVE and the unbounded bulk space. Mapping between the two spaces leads to a robust implementation of flux boundary condition along with an instantaneous tracking of the individual loops throughout the history of deformation. The formulation considers glide planes in the RVE as discrete mathematical objects. These glide planes can be shifted and stitched together to define a superspace for the evolution of dislocation loops in the infinite crystal. For such a construct, the RVE will serve not only as a continuum RVE but also as a lattice RVE. That is, it can be thought of as a supercell of the crystal lattice. The implementation has been carried out using the DDD code MoDELib (Po et al., 2014). Some test cases for the periodic loop kinematics are presented for testing the accuracy of the proposed scheme. The stress implementation part will follow in future part of this work.

## 2. Representative volume element in discrete dislocation dynamics

Investigating the macroscopic response of a bulk crystal using microscopic dislocation fields requires introducing a sub-volume of the macroscopic crystal called an RVE and identifying invariant fields of interest based on the RVE introduced. Dislocation density field and elastic stress field are the fundamental fields for setting up the RVE problem in DDD. While the former field describes incompatibility in crystal lattice storing details about geometry of distortions, the latter field provides driving force for evolution of dislocation microstructure. At this point, it is essential to distinguish between the macroscopic and microscopic response of the crystals in terms of the invariant fields. Gradients of the relevant fields are significant at the microscopic scale as compared to the macroscopic scale. This implies that dislocation arrangements attributed to their line direction and burgers vector is deterministic at the microscale as opposed to their ensemble averages at the macroscale (Kröner, 2001).

Consider a macroscopically large crystal V with boundary $\partial V$ having linear dimensions on the order of D as shown in Figure 1. Let the macroscopic volume be subjected to a remote traction $\mathbf{T}_0$ on its boundary such that the elastic, plastic and dislocation density fields in the crystal are uniform at the macroscopic scale. Let us consider a sub-volume, $\Omega$ with boundary $\partial \Omega$, of the bulk crystal with linear dimensions, d, that is sufficiently small compared with the macroscopic crystal, i.e., d/D<<1, but large compared to a suitable measure of the inter-dislocation spacing and it is at least a distance d inside the edges of the macroscopic volume. Let us call this region of the macroscopic volume where the RVE can reside as interior region of the macroscopic volume. For the sub-volume $\Omega$ to be an RVE of the bulk crystal, three conditions must be satisfied, as explained below. Before diving into the constraints, however, average field quantities and the notion of statistical homogeneity must be introduced. The average of an arbitrary field, $\mathbf{F}$, over a volume of interest, V, is denoted by $\bar{\mathbf{F}}_V$ and is defined as

$$\bar{\mathbf{F}}_V = \frac{1}{V} \int_V \mathbf{F} \, dV. \tag{1}$$

For discussions about the statistical homogeneity, let us consider another sub-volume, $\Omega_0$, in the interior region of macroscopic volume, with boundary $\partial \Omega_0$ having same size and shape as that of $\Omega$ such that the centroid of $\Omega_0$ is located at the origin. Let us consider that all points $\mathbf{z} \in \Omega_0$ can be translated by a constant translational vector $\mathbf{y}$ so as to get all points $\mathbf{x} \in \Omega$ of the RVE (Figure 1), i.e., $\mathbf{x} = \mathbf{y} + \mathbf{z}$, and the centroid of $\Omega$ is at $\mathbf{y}$. The field variable $\mathbf{F}$ in $\Omega$ can be written as

$$\mathbf{F} = \mathbf{F}(\mathbf{x}) = \mathbf{F}(\mathbf{y} + \mathbf{z}). \quad (2)$$

Average of the field, **F**, over the RVE can be written as

$$\bar{\mathbf{F}}(\mathbf{y}) = \frac{1}{\Omega} \int_\Omega \mathbf{F}(\mathbf{x}) \, d\Omega = \frac{1}{\Omega_0} \int_{\Omega_0} \mathbf{F}(\mathbf{y} + \mathbf{z}) \, d\Omega_0. \quad (3)$$

Thus, $\bar{\mathbf{F}}(\mathbf{y})$ specifies the moving average of the field in the macroscopic volume (Nemat-Nasser and Hori, 2013). Corresponding to a constant remote field, requirement for statistical homogeneity of the macroscopic volume entails that the moving averages of the fields of interest taken over the sufficiently large sub volumes is independent of shape, size and location of the centroid of the sub-volumes. Thus, the averages of the fields taken over the RVE is same as that of the interior regions of the macroscopic volume which is statistically the same as the average fields in the macroscopic volume. Let us consider the constant remote field to be $\mathbf{F}^0$, the average fields in the RVE can be written as

$$\bar{\mathbf{F}}(\mathbf{y}) \approx \mathbf{F}^0. \quad (4)$$

With the definition of the average quantities and the statistical homogeneity, we now move towards defining the necessary constraints for the RVE.

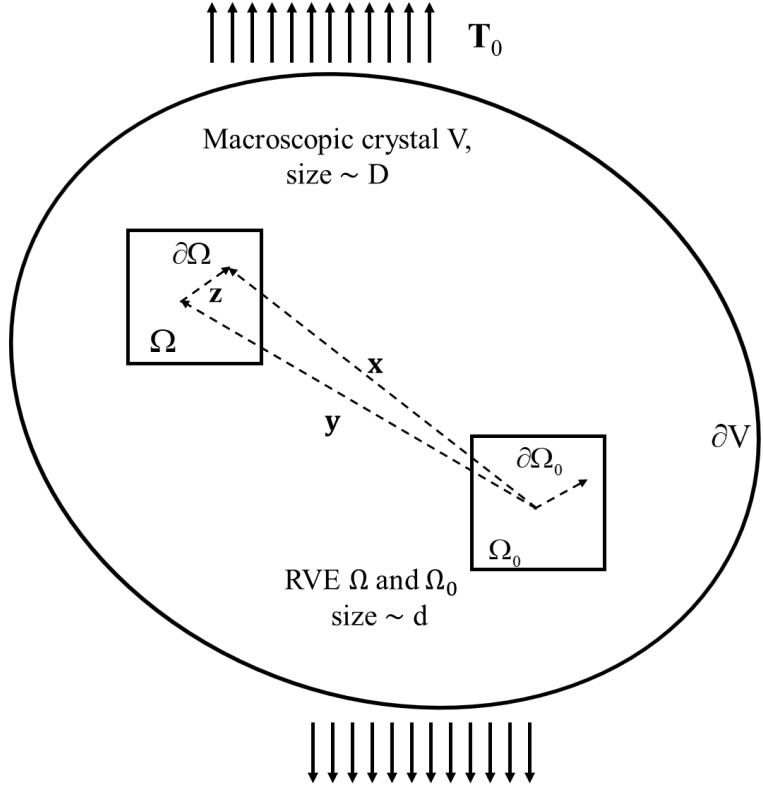

Figure 1: A schematic for a bulk crystal with arbitrary remote traction boundary condition. $\Omega$ and $\Omega_0$ are two equivalent REVs.

### 2.1. Mechanical field constraints

The mechanical field constraints establish the equivalence between the RVE volume and the bulk crystal in terms of average mechanical fields. Let us first consider the stress field. The (total) local stress field, $\boldsymbol{\sigma}^\text{T}$, in the crystal containing dislocations is

$$\boldsymbol{\sigma}^\text{T} = \boldsymbol{\sigma}^0 + \boldsymbol{\sigma}, \tag{5}$$

where $\boldsymbol{\sigma}^0$ represents the stress field due to remote traction, $\mathbf{T_0}$, and $\boldsymbol{\sigma}$ represents the internal elastic stress field in the crystal due to dislocations. The average stress field in the crystal is

$$\bar{\boldsymbol{\sigma}}^\text{T}_\text{V} = \bar{\boldsymbol{\sigma}}^0_\text{V} + \bar{\boldsymbol{\sigma}}_\text{V}. \tag{6}$$

By virtue of Albenga's theorem (El-Azab, 2000), the average internal elastic stress field in the crystal vanishes identically for a volume under static equilibrium, i.e. $\bar{\boldsymbol{\sigma}}_\Omega = \bar{\boldsymbol{\sigma}}_V = \mathbf{0}$. Hence, the average total stress field in the crystal becomes

$$\bar{\boldsymbol{\sigma}}_V^T = \bar{\boldsymbol{\sigma}}_V^0 = \frac{1}{V}\oint_{\partial V} \mathbf{x} \otimes \mathbf{T}_0 \, dS, \tag{7}$$

where $\mathbf{x}$ represents boundary coordinate of V. Since $\bar{\boldsymbol{\sigma}}_\Omega^0 = \bar{\boldsymbol{\sigma}}_V^0$, the average stress over $\Omega$ must be the same as the average stress in the bulk crystal for $\Omega$ to be an RVE; that is

$$\bar{\boldsymbol{\sigma}}_\Omega^T = \bar{\boldsymbol{\sigma}}_V^T. \tag{8}$$

This mechanical constraint is the basis of the stress-field calculation in the RVE, which will be the subject of the follow-up communication. For now, we turn our attention towards the kinematic aspect of plastic deformation of a crystal. Denote the local distortion of the crystal by $\boldsymbol{\beta}^T$. This distortion can be split into a plastic part due to dislocation movement, $\boldsymbol{\beta}^p$, an elastic part associated with the remote load, $\boldsymbol{\beta}^0$, and an elastic part due to the dislocations, $\boldsymbol{\beta}^e$ (El-Azab, 2000; Kossecka and Dewit, 1977; Kröner, 1981; Mura, 1987),

$$\boldsymbol{\beta}^T = \boldsymbol{\beta}^0 + \boldsymbol{\beta}^e + \boldsymbol{\beta}^p. \tag{9}$$

We consequently can write the average total distortion of the crystal $(\bar{\boldsymbol{\beta}}_V^T)$, as additive decomposition of average elastic distortion due to remote load $(\bar{\boldsymbol{\beta}}_V^0)$, average elastic $(\bar{\boldsymbol{\beta}}_V^e)$ and plastic $(\bar{\boldsymbol{\beta}}_V^p)$ distortions due to dislocations; that is

$$\bar{\boldsymbol{\beta}}_V^T = \bar{\boldsymbol{\beta}}_V^0 + \bar{\boldsymbol{\beta}}_V^e + \bar{\boldsymbol{\beta}}_V^p. \tag{10}$$

$\bar{\boldsymbol{\beta}}_V^0$ and $\bar{\boldsymbol{\beta}}_V^e$ are related with $\bar{\boldsymbol{\sigma}}_V^0$ and $\bar{\boldsymbol{\sigma}}_V^e$ via Hooke's law of linear elasticity; that is

$$\bar{\boldsymbol{\sigma}}_V^0 = \mathbf{C} : \bar{\boldsymbol{\beta}}_V^0 \tag{11}$$

and

$$\overline{\boldsymbol{\sigma}}_V^e = \mathbf{C} : \overline{\boldsymbol{\beta}}_V^e = \mathbf{0}, \tag{12}$$

where **C** is a fourth-order elastic stiffness tensor. From Eq. (11), we immediately see that

$$\overline{\boldsymbol{\beta}}_\Omega^0 = \overline{\boldsymbol{\beta}}_V^0. \tag{13}$$

Additionally, for any arbitrary volume V, we can write $\overline{\boldsymbol{\beta}}_V^e = \mathbf{0}$ as is evident from Eq. (12). For the RVE, this constraint for vanishing average elastic distortion can be written as

$$\overline{\boldsymbol{\beta}}_\Omega^e = \overline{\boldsymbol{\beta}}_V^e = \mathbf{0}. \tag{14}$$

Since the total distortion is a compatible field, we can write the total distortion as gradient of displacement and use the condition for statistical homogeneity (Eq. (4)) to write

$$\overline{\boldsymbol{\beta}}_\Omega^T = \frac{1}{\Omega} \int_\Omega \nabla \mathbf{u}^T d\Omega = \frac{1}{\Omega} \oint_{\partial \Omega} \mathbf{u}^T \otimes d\mathbf{A} = \overline{\boldsymbol{\beta}}_V^T. \tag{15}$$

This implies that the average plastic distortion in the RVE must be same as that of the bulk crystal; that is

$$\overline{\boldsymbol{\beta}}_\Omega^p = \overline{\boldsymbol{\beta}}_V^p. \tag{16}$$

Eq. (16) can also be casted in terms of rate of plastic distortion. The volume averaged rate of plastic distortion in the crystal RVE must be equal to the volume averaged rate of plastic distortion in the bulk crystal. Mathematically, the equation can be written as

$$\overline{\dot{\boldsymbol{\beta}}}_\Omega^p = \overline{\dot{\boldsymbol{\beta}}}_V^p. \tag{17}$$

The constraint given in Eq. (16) and (17) will be useful in deriving the flux boundary conditions which is discussed next.

## 2.2. Microstructure constraint

The microstructural constraint establishes the equivalence between the RVE and the bulk crystal in terms of the dislocation density field. The plastic distortion associated with a dislocation loop can be represented with a surface integral spanning the dislocation loop (Eq. (18)). In Eq. (18), $\delta$ is the Dirac delta function and $\delta(\mathbf{x} - \mathbf{x}')$ indicates that the plastic distortion is confined by the dislocation loop, $\mathbf{b}$ represents the Burgers vector and $d\mathbf{A}$ represents infinitesimal oriented area.

$$\boldsymbol{\beta}^p = \int_S \delta(\mathbf{x} - \mathbf{x}') \, d\mathbf{A} \otimes \mathbf{b}. \tag{18}$$

The crystal undergoing plastic deformation experiences a state of self-stress (Mura, 1987), which depends on the curl of the plastic distortion tensor, $\nabla \times \boldsymbol{\beta}^p$. The curl of plastic distortion is also known as the dislocation density tensor, $\boldsymbol{\alpha}$, and is represented by

$$\boldsymbol{\alpha} = \oint_L \delta(\mathbf{x} - \mathbf{x}') \, d\mathbf{L} \otimes \mathbf{b}, \tag{19}$$

where $d\mathbf{L}$ represents an infinitesimal differential dislocation line element. Eq. (19) can be interpreted in the following way: the component of the dislocation density tensor, $\alpha_{ij}$, represents a dislocation line that intersects a unit normal area in the $x_i$-direction having an $x_j$ component of the Burgers vector. The divergence of the dislocation density tensor must vanish, i.e., $\nabla \cdot \boldsymbol{\alpha} = \mathbf{0}$ (Kröner, 1981). This constraint on the dislocation density tensor implies that a dislocation cannot end inside the crystal. Hence, dislocations must exist either as closed loops, participate in network as per Frank's rule, or end at the surfaces of the crystal leaving behind slip traces. Since we are representing a macroscopically large crystal as a smaller sub-volume using RVE, the boundaries of the RVE represent interfaces to the periodically repeating sub-volume mimicking the macroscopically large crystal. Hence, the dislocations cannot end at the boundaries of the RVE and the dislocation population must be present as a system of closed loops in bulk space. For such a case, the volume averaged dislocation density tensor must vanish. Mathematically, this can be represented as

$$\bar{\boldsymbol{\alpha}}_V = \frac{1}{V}\int_V \boldsymbol{\alpha}\, dV = \frac{1}{V}\int_V dV \oint_L \delta(\mathbf{x}-\mathbf{x}')\, d\mathbf{L} \otimes \mathbf{b} = \frac{1}{V}\oint_L d\mathbf{L} \otimes \mathbf{b} = \mathbf{0}, \tag{20}$$

with L being the set of all dislocation lines in the crystal volume V. The last equation implies that the crystal must be undergoing a uniform (curl-free) plastic distortion at the macroscale and the condition for statistical homogeneity requires

$$\bar{\boldsymbol{\alpha}}_\Omega = \bar{\boldsymbol{\alpha}}_V = \mathbf{0}. \tag{21}$$

Resolving dislocation density on individual slip systems, the average dislocation density can be written as:

$$\frac{1}{V}\sum_i \sum_{L_i} \oint_{L_i} d\mathbf{L} \otimes \mathbf{b}^i = \mathbf{0}, \tag{22}$$

where, $\sum_i$ represents summation over individual slip systems i and $\sum_{L_i}$ indicates the summation over individual dislocation loops within $i^{th}$ slip system. Since all slip systems may not be active during the course of deformation of the crystal, Eq. (22) can be invoked for individual slip systems as well. Correspondingly, the average dislocation density tensor, $\bar{\boldsymbol{\alpha}}_\Omega$, in the RVE can be written using an open line integral as the dislocation loops exiting the boundaries of the RVE may not be closed. That is,

$$\bar{\boldsymbol{\alpha}}_\Omega = \frac{1}{\Omega}\sum_i \sum_{L_i} \int_{L_i} d\mathbf{L} \otimes \mathbf{b}^i = \sum_i \sum_{L_i}(\mathbf{F}-\mathbf{S}) \otimes \mathbf{b}^i = \mathbf{0}, \tag{23}$$

where **F** and **S** corresponds to the end point positions of dislocation loop, $L_i$, exiting the boundary of the RVE. The fact that dislocations may be participating junctions and form networks inside the RVE does not change the above arguments. These points are also shown in Figure 2 where the microstructural constraints on the dislocation loops in an RVE are presented. Eq. (23) provides the constraint for ensuring that the dislocation population exist as closed loops in the RVE and requires that the dislocation density must be balanced across the boundaries of the RVE. We see that for every dislocation segment belonging to a crystallographic family and exiting a boundary of the RVE, there must be another dislocation segment belonging to the same family entering the RVE from a periodically equivalent boundary. This ensures that $\bar{\boldsymbol{\alpha}}_\Omega$ remains zero for the RVE. Hence, the boundaries of the crystal RVEs are present just to provide numerically tractable solutions and should not play any role in influencing the dynamics of dislocations.

Eq. (21) can also be casted in terms of volume averaged rate of dislocation density tensor as

$$\bar{\boldsymbol{\alpha}}_\Omega = \bar{\boldsymbol{\alpha}}_V = \mathbf{0}. \tag{24}$$

Kosevich, (1965) wrote the rate of dislocation density tensor in terms of the rate of plastic distortion tensor as

$$\dot{\boldsymbol{\alpha}} + \nabla \times \dot{\boldsymbol{\beta}}_p = \mathbf{0}, \tag{25}$$

where the rate of plastic distortion tensor, $\dot{\boldsymbol{\beta}}_p$, for closed dislocation loops is given by

$$\dot{\boldsymbol{\beta}}_p = \sum_i \sum_{L_i} \oint_{L_i} \delta(\mathbf{x} - \mathbf{x}')(d\mathbf{L} \times \mathbf{v}) \otimes \mathbf{b}^i, \tag{26}$$

where $\mathbf{v}$ is the dislocation velocity of the differential dislocation line element, $d\mathbf{L}$. In terms of the volume averaged rate of dislocation density tensor in the RVE, Eq. (25) can be written as

$$\frac{1}{\Omega}\int_\Omega \dot{\boldsymbol{\alpha}} \, d\Omega + \frac{1}{\Omega}\int_\Omega \nabla \times \dot{\boldsymbol{\beta}}_p \, d\Omega = \mathbf{0}. \tag{27}$$

Using Eq. (24), Eq. (27) becomes

$$\int_\Omega \nabla \times \dot{\boldsymbol{\beta}}_p \, d\Omega = \mathbf{0}. \tag{28}$$

This implies that

$$\int_\Omega \nabla \times \dot{\boldsymbol{\beta}}_p \, d\Omega = \sum_i \sum_{L_i} \int_\Omega d\Omega \oint_{L_i} \nabla \times [\delta(\mathbf{x} - \mathbf{x}')(d\mathbf{L} \times \mathbf{v}) \otimes \mathbf{b}^i] = \mathbf{0}, \tag{29}$$

which after some algebraic manipulations can be written in the form

$$\sum_i \sum_{L_i} \int_\Omega d\Omega \oint_{L_i} \left[\left(\nabla \cdot (\delta(\mathbf{x} - \mathbf{x}') v\hat{\boldsymbol{\xi}})\right)(d\mathbf{L} \otimes \mathbf{b}^i) - (d\mathbf{L} \cdot \nabla(\delta(\mathbf{x} - \mathbf{x}') v))(\hat{\boldsymbol{\xi}} \otimes \mathbf{b}^i)\right] = \mathbf{0}, \tag{30}$$

where $\mathbf{v} = v\hat{\boldsymbol{\xi}}$ and $\hat{\boldsymbol{\xi}}$ is a unit vector along $\mathbf{v}$. Since there does not exist any flux of dislocations along its own line direction, the second expression on the left-hand side becomes zero, i.e. $(d\mathbf{L} \cdot \nabla(\delta(\mathbf{x} - \mathbf{x}') v))(\hat{\boldsymbol{\xi}} \otimes \mathbf{b}^i) = \mathbf{0}$. Eq. (*30*) becomes

$$\sum_i \sum_{L_i} \int_\Omega d\Omega \oint_{L_i} \left[\left(\nabla \cdot (\delta(\mathbf{x} - \mathbf{x}') v\hat{\boldsymbol{\xi}})\right)(d\mathbf{L} \otimes \mathbf{b}^i)\right] = \mathbf{0}. \tag{31}$$

Using divergence theorem, we can write

$$\sum_i \sum_{L_i} \int_{\partial\Omega} dS \oint_{L_i} \left[(\delta(\mathbf{x} - \mathbf{x}') \mathbf{v} \cdot \mathbf{n})(d\mathbf{L} \otimes \mathbf{b}^i)\right] = \mathbf{0}, \tag{32}$$

where $\mathbf{n}$ is the outward normal to the boundary of the RVE volume, $\Omega$. Eq. (32) indicates that the normal dislocation flux vanishes identically for closed dislocation loops on any internal surface of the macroscopic volume, i.e., the rate of the slip on any internal surface of the macroscopic volume is equally contributed by dislocation moving in the opposite directions (El-Azab, 2000). Now, Eq. (32) is needed to be written as an open line integral due to dislocations exiting the boundaries of the RVE, i.e.,

$$\sum_i \sum_{L_i} \int_{\partial\Omega} dS \int_{L_i} \left[(\delta(\mathbf{x} - \mathbf{x}') \mathbf{v} \cdot \mathbf{n})(d\mathbf{L} \otimes \mathbf{b}^i)\right] = \mathbf{0}, \tag{33}$$

which for a surface of an arbitrary RVE becomes

$$\sum_i \sum_{L_i} \int_{L_i} \left[(\delta(\mathbf{x} - \mathbf{x}') \mathbf{v} \cdot \mathbf{n})(d\mathbf{L} \otimes \mathbf{b}^i)\right] = \mathbf{0}. \tag{34}$$

Eq. (34) implies that for each straight dislocation segment, **L**, exiting a boundary of the RVE, $\partial\Omega$, with a scalar velocity $v_n$ projected normal to the boundary, **n**, there must be another dislocation segment, $\mathbf{L}' = \mathbf{L}$, entering the RVE from a periodically equivalent boundary having a scalar velocity $v_{n'}$ projected normal to the periodically equivalent boundary, $\mathbf{n}'$, with $\mathbf{n} + \mathbf{n}' = \mathbf{0}$. Eq. (34) provides a more rigorous constraint than Eq. (23) on the dislocations exiting the boundaries of the RVE and requires that the total outward flux must be equal to the total inward flux at any instant of time during the deformation of the crystal. Eq. (34) requires tracking the swept area by dislocation loops and balancing the area of the loops exiting the RVE. Additionally, this constraint automatically ensures that the dislocation density is balanced at the boundaries. Eqs. (23) and (34) together constitute the necessary requirements on the dislocation flux for the crystal to be undergoing uniform plastic distortion.

In Figure 2, we present schematics of the bulk crystal as well as the RVE approximations for the bulk crystal to illustrate the kinematic microstructural constraints on dislocation loops. As is evident in Figure 2(a), the sub-volume, $\Omega$, is not an RVE of the bulk crystal, V, because the constraints given by Eqs. (23) and (34) are not satisfied. In Figure 2 (b), the dislocation density is balanced across the boundary of the RVE but the velocities of the dislocation segments at the boundaries of the RVE do not match. Hence, Eq. (34) is not satisfied and the sub-volume cannot be considered as an RVE. Lastly, Figure 2 (c) represents a valid RVE as both Eqs. (23) and (34) are satisfied.

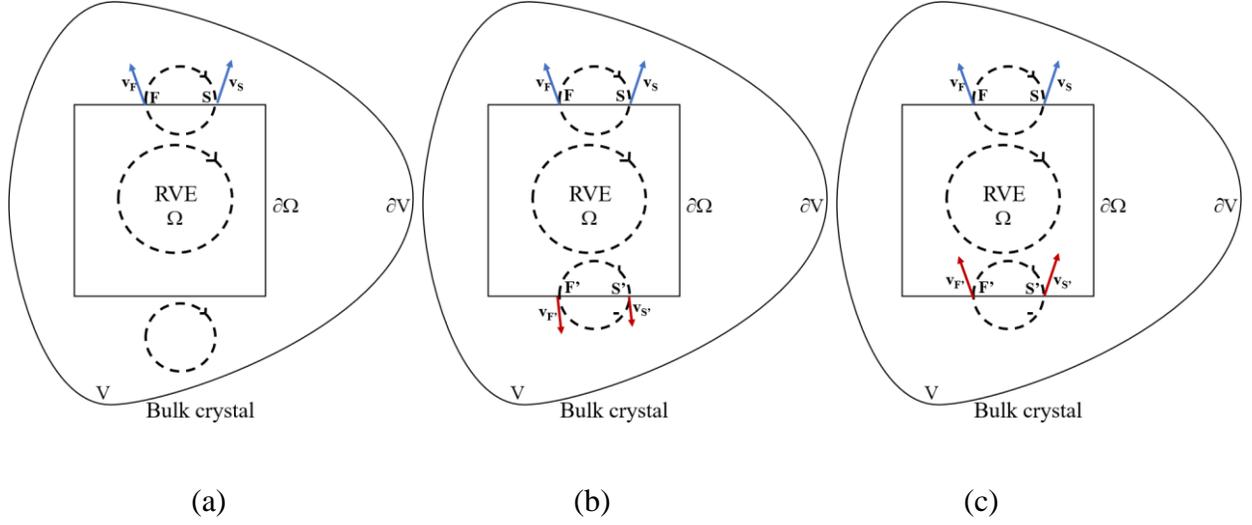

Figure 2: A schematic for modeling an RVE as a sub-volume of bulk material for satisfying the kinematic constraints on the RVE. (a) An invalid RVE representation for the bulk crystal (Eqs. (23) and (34) are not satisfied). (b) An invalid RVE representation for the bulk crystal (Eq. (34) is not satisfied). (c) A valid RVE representation with $\mathbf{v}_F = \mathbf{v}_{F'}$ and $\mathbf{v}_S = \mathbf{v}_{S'}$ (The arrow at the points show the direction of velocities associated with boundary points $\mathbf{F}$ and $\mathbf{S}$).

### 2.3. Lattice constraint

To balance dislocation flux across the boundaries of the RVE, the RVE domain must be commensurate with the unit cell of the material under study. This requires that the RVE under consideration must be a supercell of the crystal lattice. This condition ensures that glide planes belonging to same crystallographic families exist on periodically equivalent faces of the RVE. Thus, the lattice constraint allows for reinsertion of dislocation segments exiting the RVE on consistent crystallographic planes.

Sections 2.1-2.3 provide the necessary RVE constraints for modeling bulk deformation of crystals via DDD. In contrast to the widely used minimum image convention approach to preserve the dislocation flux continuity across the RVE boundaries, in this paper we develop a novel *topological* method that satisfies the RVE *kinematic* constraints for dislocation flux boundary conditions (Eqs. (23) and (34)) identically and is well suited to impose the stress RVE constraint (Eq. (8)). Our implementation is restricted to dislocation glide, where crystal dislocations are confined to move on specific planes of the crystal lattice. Implementation of the stress RVE constraint along with this topological implementation of flux boundary conditions will be the subject of a follow-up work.

## 3. Dual topology for periodic loop kinematics

### 3.1. Glide planes in the RVE superlattice

The use of periodic flux boundary condition may introduce numerical artifacts by altering the internal stress state of the crystal. This happens due to spurious annihilations upon periodic reentry of dislocation segments which inadvertently results in reduction of dislocation density. Hence, it is essential to address self-annihilations from the viewpoint of RVE kinematics.

Here, we provide a robust framework for choosing the size of the RVE such that the self-annihilation distance could be increased while meeting the RVE requirements (see also Madec et al. (2004)). The self-annihilation distance is the distance which a dislocation loop travels before it interacts with its own replica due to periodic reentry, hence resulting annihilation of a part of the loop. RVE flux boundary condition requires that a lattice plane of the same crystallographic family must exist on periodically equivalent RVE faces. Dislocation segments exiting the RVE from a face can be reinserted into the RVE from its periodically equivalent face. Two faces of the RVE with $\mathbf{n}$ and $\mathbf{n}'$ as outward normal to the faces are periodically equivalent if $\mathbf{n} + \mathbf{n}' = \mathbf{0}$. This implies that lattice planes can be shifted by the RVE dimensions and stitched together to give rise to planes in the bulk crystal space, henceforth called the bulk crystal planes (or bulk glide planes). These planes extend outside of the RVE and can be mapped to the RVE through shift vectors or modulo periodicity vectors. Hence, there exists a mapping between the bulk crystal space and the RVE space. To define the mapping, we consider the RVE, $\Omega$, with boundary, $\partial\Omega$, representative of a bulk crystal V. Let $\aleph$ represents the mapping between the bulk crystal space and the RVE space, i.e., $\aleph: V \to \Omega$,

$$\mathbf{X} = \{\aleph(\mathbf{x}) \mid \mathbf{x} \in R^3, \mathbf{X} \in R'^3\}. \tag{35}$$

where $R^3$ represents the three-dimensional Euclidean space and $R'^3 \subset R^3$ is the set of all points confined by the RVE. Eq. (35) simply states that points in the Euclidean space $R^3$ can be mapped to a point in the RVE, $R'^3$. We are interested in points which reside on lattice planes. We would like to utilize this mapping for using periodically equivalent crystallographic lattice planes to define bulk glide planes. A crystallographic lattice plane can be defined with a position $\mathbf{x_0}$ and a normal $\hat{\mathbf{m}}$ such that $\hat{\mathbf{m}} \cdot (\mathbf{x} - \mathbf{x_0}) = 0$. With respect to the origin of the crystal RVE, the plane height will be $\hat{\mathbf{m}} \cdot \mathbf{x_0}$. Similarly, periodically equivalent (crystallographic) lattice planes are the crystallographic lattice planes where dislocation segments can move during the course of RVE deformation including those planes where the segments can move after their periodic reentry into the RVE. Now, the lattice constraint for two periodically equivalent lattice planes defined by the plane normal $\hat{\mathbf{m}}$ and lattice points $\mathbf{x_1}$ and $\mathbf{x_2}$ on the two periodically equivalent lattice planes can be written as

$$\hat{\mathbf{m}} \cdot (\mathbf{x_1} - \mathbf{x_2} + \mathbf{B}\boldsymbol{\xi}) = 0 \qquad \text{for some integer vector } \boldsymbol{\xi}, \tag{36}$$

where $\mathbf{B} = \mathbf{AN}$ represents the matrix of primitive vectors of the RVE superlattice, $\boldsymbol{\xi}$ is an integer translational vector representing the images of the RVE, the matrix $\mathbf{A}$ represents the basis of the crystal lattice along columns and $\mathbf{N}$ is an integer matrix. The two lattice points $\mathbf{x_1}$ and $\mathbf{x_2}$ are separated by translational vector $\mathbf{B}\boldsymbol{\xi}$ and can be represented using $\mathbf{A}$ as $\mathbf{x_1} = \mathbf{A}\boldsymbol{\xi_1}$ and $\mathbf{x_2} = \mathbf{A}\boldsymbol{\xi_2}$ for integers $\boldsymbol{\xi_1}$ and $\boldsymbol{\xi_2}$. The plane normal $\hat{\mathbf{m}}$ can be written as $\hat{\mathbf{m}} = \frac{\mathbf{A}^{-T}\boldsymbol{\zeta}}{\|\mathbf{A}^{-T}\boldsymbol{\zeta}\|}$ with $\boldsymbol{\zeta}$ being an integer vector which represents the plane normal in the reciprocal space. The total number of parallel planes belonging to same crystallographic family between two periodically equivalent planes is (see Appendix for detailed derivation)

$$n_p(\boldsymbol{\xi}, \mathbf{N}) = \text{GCD}(\boldsymbol{\zeta}^T \mathbf{N}), \qquad (37)$$

where GCD is the greatest common divisor of the elements of the integer vector $\boldsymbol{\zeta}^T \mathbf{N}$. Given the interplanar distance for the family of planes at hand, $d_{min}$, the distance between periodically equivalent planes $d_p$ is given by

$$d_p(\boldsymbol{\zeta}, \mathbf{N}) = d_{min} n_p. \qquad (38)$$

Eq. (37) provides a quantitative measure with which the total number of crystallographic lattice planes that are available for the dislocations to glide, such that the microstructural and the lattice constraints on the dislocation loops are satisfied, can be calculated (see Appendix for more details).

Given a commensurate superlattice for the lattice of the material under study, Eq. (36) gives the shift vectors between the periodically equivalent lattice planes with respect to the dimensions of the RVE. Hence, periodically equivalent planes in RVE can be shifted with respect to a reference plane to construct the bulk glide planes as demonstrated in Figure 3. The integer shift vectors, $\boldsymbol{\xi}$, are also indicated in the figure. A trivial choice for the reference plane is the glide plane of a loop. Eq. (36) also gives a measure of the distance that segments of a dislocation loops travel under pure expansion before they self-annihilate. Mathematically, the condition for the self-annihilation of a dislocation loop is given by Eq. (39) and can be represented as a special case of Eq. (36) when the two lattice points representing two periodically equivalent lattice planes become same (see Appendix).

$$\boldsymbol{\zeta}^T(\mathbf{N}\boldsymbol{\xi}) = 0. \tag{39}$$

Eq. (39) is a Diophantine equation and admits only integer solution for $\boldsymbol{\xi}$. Thus, a solution of Eq. (39) provides the shift vectors under uniform expansion of a dislocation loop before the segments associated with the loop will undergo self-annihilation. Hence, size of the RVE can be chosen such that the dislocation mean free path before self-annihilation can be maximized.

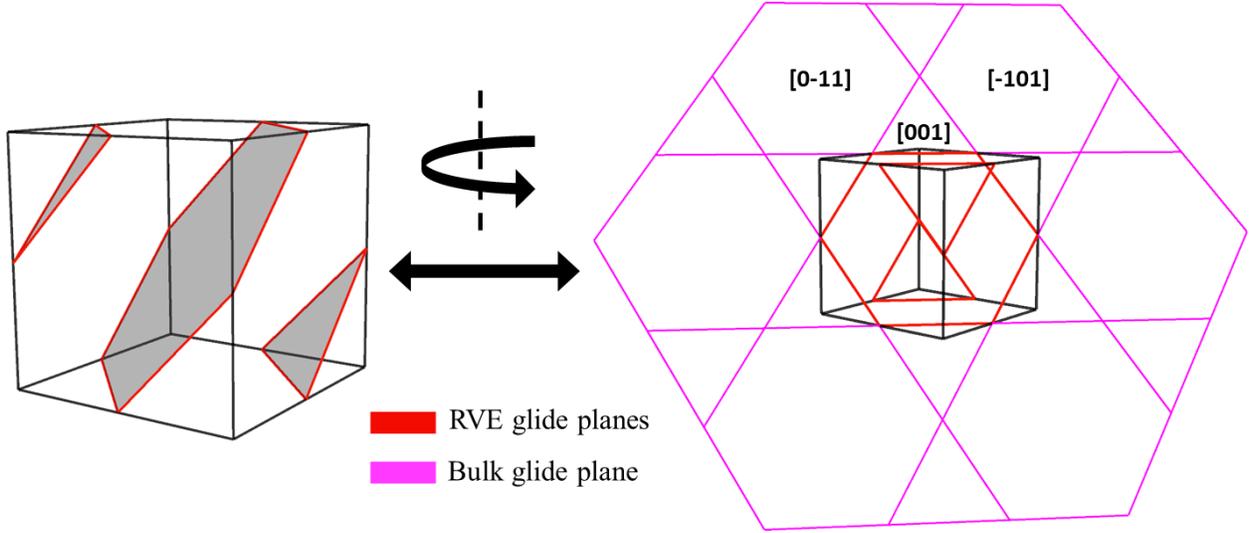

Figure 3: A schematic indicating the mapping between the RVE glide planes and bulk glide plane. The integer shift vectors ($\boldsymbol{\xi}$ in Eq. (36)) are also marked in the shifted glide planes of the RVE to illustrate the mapping presented in Eq. (35). Glide planes are shaded in the left figure. The RVE is rotated to highlight the bulk glide plane in the image on the right. Mapping between the glide planes and the bulk glide planes through translational vector, $\boldsymbol{\xi}$, are shown on the right.

## 3.2. Periodic representation of dislocation loops

In addition to dislocation transport, dislocation density evolves with the plastic deformation of a crystal via discrete events such as glissile junction formation and cross slip, giving rise to a complicated arrangement of dislocation networks. Dislocation segments involved in these glide events always leave behind a network consisting of zero Burgers vector segments, henceforth called the virtual dislocation density. The virtual dislocation densities, as opposed to glide dislocation densities, do not contribute to the internal stress fields of the crystal. The sole purpose of virtual densities is to realize closure of dislocation loops, one of the necessary kinematic requirements for the RVE. The kinematic microstructural constraint requires that the dislocation flux must be balanced across the RVE boundaries while preserving the dislocation networks consisting of glide and virtual densities. These discrete glide events make enforcing the kinematic microstructural constraints very difficult. To address this issue, we provide a dual topological representation of the dislocations: one representation for the dislocation loops and the other for the dislocation networks. This allows for tracking the evolution of loops separately from the evolution of the network. This ensues defining two separate yet mapped spaces (Eq. (35)) of a crystal namely: a crystal RVE space where the dislocations network governs the deformation dynamics of the RVE and a bulk crystal space of the RVE where the dislocation loops evolve with time. The dislocation loops are topologically constrained to move on bulk glide planes. The dislocation loops are discretized into segments, henceforth called the loop segments (or dislocation loop segments). These dislocation loop segments can be mapped into the crystal RVE through translational vectors. The mapped segments, henceforth called the network segments (or dislocation network segments), can move in the crystal RVE. The network segments are topologically constrained to move on the glide planes within the RVE. They can participate in discrete events such as junction formation and cross slip. Two or more network segments in the RVE can form junctions resulting in a network segment consisting of multiple dislocation loop segments in the bulk crystal space. A cross slip event in the RVE can also be accounted using a similar manner wherein a perfect screw network segment can cross slip into a conjugate plane by forming a new loop in the conjugate plane. The new loop is formed in the bulk crystal space and the network segments corresponding to the loop are mapped into the RVE through translational vectors. In such a case, the cross-slipping segment will leave behind a zero Burgers vector segment (virtual density) at the cross-slipping site. Motion of the dislocation segments can be realized by solving for the velocity of the dislocation network nodes. Corresponding inverse mapping with the bulk crystal space modifies the dislocation loops to reflect the glide and transport events in the crystal RVE. A realization of the adopted dual topology is presented in Figure 4 where a loop structure as well as the network structure has been highlighted for a dislocation loop in Figure 4(a) and for two dislocation loops forming a junction in Figure 4(b).

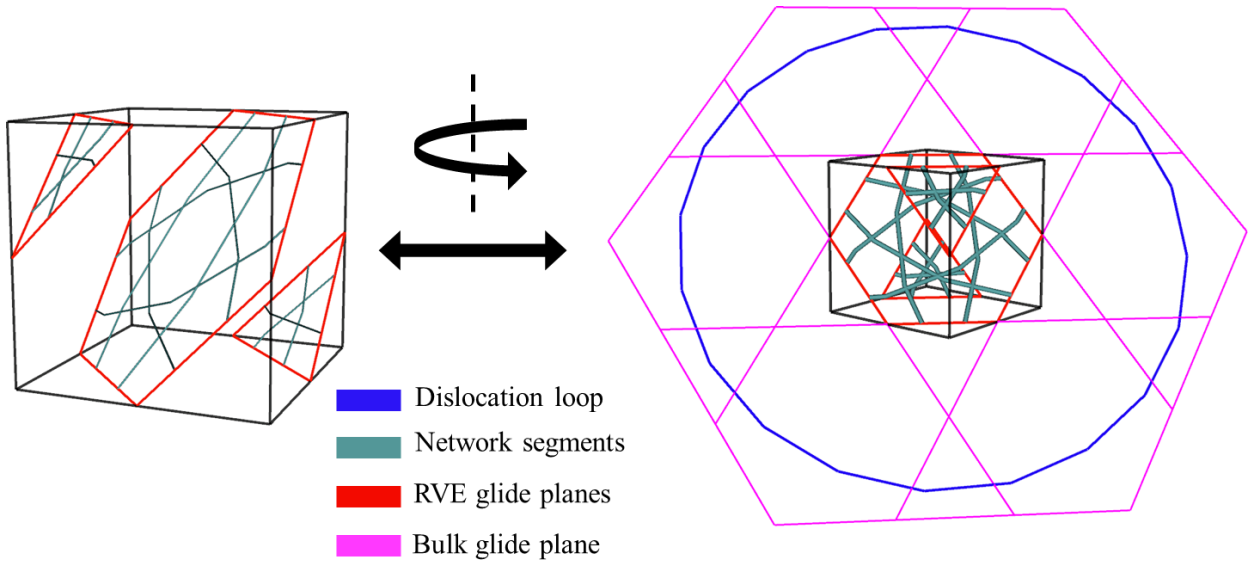

(a)

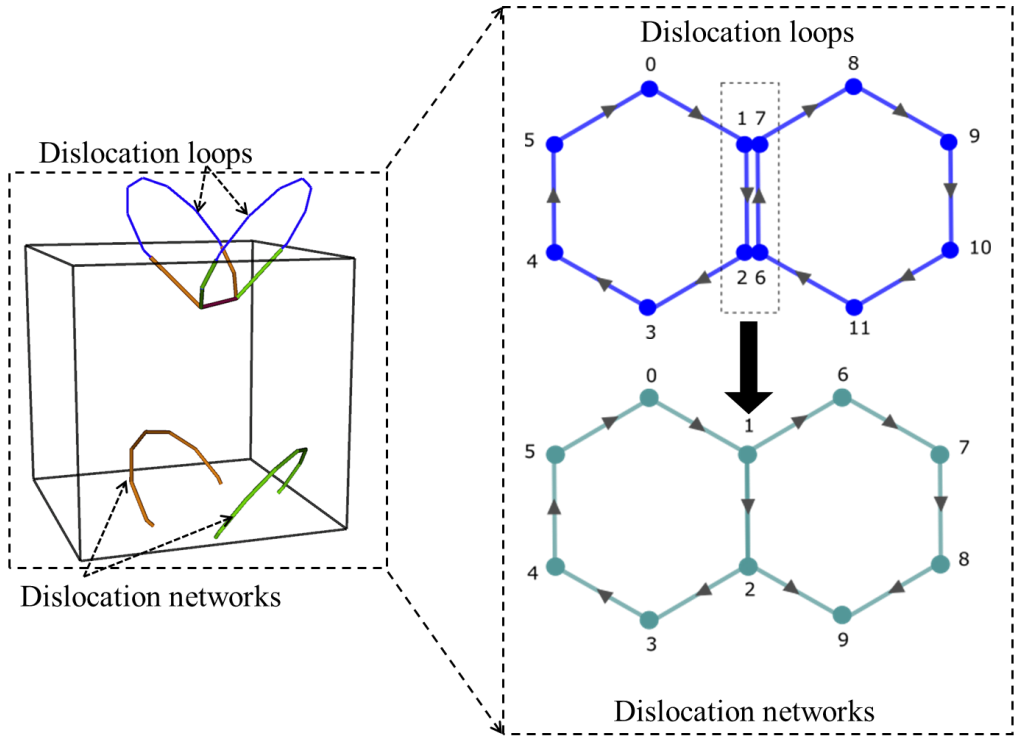

(b)

Figure 4: Dual topological representation of dislocation loops and dislocation networks (a) Representation of topological entities namely dislocation loop, dislocation network segments, glide planes, and bulk glide plane in the dual space. Dislocation network segments are constrained to RVE glide planes as shown in the image on the left. Dislocation loop is constrained to bulk glide plane and the loop segments can be mapped to the network segments through translational vectors as shown in the image on the right. The RVE in the left image is rotated to highlight the bulk glide plane in the image on the right. (b) Representation of junction formation. The left image shows the two dislocation loops forming a junction and extending outside of the RVE whereas the network segments of the dislocation loops are mapped back into the RVE space via the translational vectors. The right image shows the schematic of the representation of the dislocation loops and the dislocation networks via the dual topological representation. The actual dynamics of the RVE is governed by the dislocation network segments in the RVE whereas loops provide a means for tracking the plastic distortion as well as its mapping with the dislocation network segments throughout the deformation of the RVE.

### 3.3. Boundary nodes in representative volume elements

As dislocation network segments exit the RVE, nodes are left at periodically equivalent boundaries, henceforth referred to as boundary nodes of the segments, for preserving the topological connectivity of dislocation segments. A schematic for the creation of boundary nodes due to a dislocation network segment exiting the RVE is shown in Figure 5 (a) and (b). These boundary nodes are constrained to move with the dislocation segments. In this work, a constrained solution for the velocity of network segment nodes has been implemented to eliminate the influence of the boundary nodes on the motion of dislocations. The nodal velocities of dislocation network segments in the RVE are updated by solving equation of motion for the dislocation network using a framework similar to Finite Element Method (FEM) (Po et al., 2014). To achieve this, the dislocation network segments are discretized with quadrature points and the equation of motion for the dislocation segments is solved using a weak formulation. More details regarding the equation of motion and its solution can be found in (Po et al., 2014). Simplistically, the final set of algebraic equations for solving the velocities of the nodes can be written as

$$\mathbf{KV} = \mathbf{F}, \tag{40}$$

where $\mathbf{K}$ represents the stiffness matrix corresponding to the dislocation network segments, $\mathbf{V}$ represents the vector of nodal velocities and $\mathbf{F}$ represents an equivalent vector derived from the quadrature point velocities. Assuming a linear interpolation of velocity between two internal nodes along a dislocation segment, let the constraint on the dislocation network nodes at the boundaries be represented using

$$\mathbf{Zv} = \mathbf{V}, \tag{41}$$

where $\mathbf{Z}$ represents the constraint matrix and $\mathbf{v}$ represents a vector of unconstrained nodal velocities. Substituting for $\mathbf{X}$ in Eq. (40), the equation becomes

$$\mathbf{KZv} = \mathbf{F}. \tag{42}$$

Hence, the solution for the nodal velocities is given as

$$\mathbf{Z}^T\mathbf{KZv} = \mathbf{Z}^T\mathbf{F}. \tag{43}$$

Now we provide a brief description for setting up the constraint matrix, $\mathbf{Z}$. Consider a dislocation network consisting of a single dislocation loop discretized with straight segments moving in the RVE as shown in Figure 5(a). At a later time-step, the network evolves, and a part of the dislocation loop exits the RVE and correspondingly the network segments are mapped back into the RVE as shown in Figure 5(b). Let us consider a straight segment AB of the dislocation loop as shown in the Figure 5(a). With the network evolution at the later time step, the segment exits the RVE boundary and network node B is mapped back into the RVE from its periodically equivalent boundary as shown in Figure 5(b). The topological connectivity for the straight segment AB corresponding the Figure 5(b) is shown in Figure 5(c). New network nodes, C and C´ are inserted at the two periodically equivalent boundaries to preserve the topology of the dislocation loops. The velocities of nodes C and C´ are constrained to the velocities of the nodes A and B. Due to the linear interpolation of velocity between the nodes A and B, the velocity constraint on C and C' can be written as: $v_C = v_{C'} = \frac{L_{BC}\, v_A + L_{AC}\, v_B}{L_{AB}}$, where $L_{AC}$, $L_{BC}$ and $L_{AB}$ are the lengths of the segments AC, BC, and AB respectively. For a dislocation ensemble consisting of multiple boundary nodes, the constraint can be written for each boundary node as described above. A global assembly is utilized for assembling the constraint equation in the constraint matrix, $\mathbf{Z}$. Subsequently the solution for the unconstrained nodal velocities has been determined. The dislocation segment nodal velocities determined from Eq. (40)-(43) are used to update the position of dislocation segment nodes. The dislocation segment nodes in turn update the position of the corresponding dislocation loop nodes in the bulk crystal space. Position of the dislocation segment nodes is updated again to reflect any changes in the glide planes of the network segment nodes due to periodic reentry. Periodic reentry results in the generation of new boundary nodes, if necessary, to preserve the topology of the dislocation network segments to be commensurate with dislocation loop segments. Hence, the loop connectivity as well as the associativity between the dual topological representation of dislocations defined by Eq. (35) is always preserved.

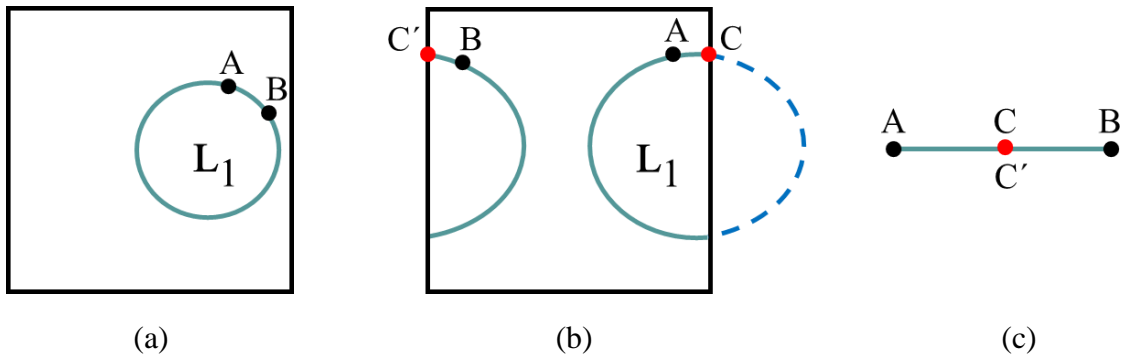

(a) (b) (c)

Figure 5: Schematic for demonstrating the creation of boundary nodes and setting of constraint matrix, **Z,** for the velocities of boundary nodes. (a) Initial configuration showing a dislocation segment, AB, of a dislocation loop, (b) evolved network configuration at a later time step showing a part of the dislocation loop mapped back into the RVE along with the network segment AB and boundary node C and C´, and (c) connectivity of the network segment AB after periodic remapping of the network segments into the RVE.

## 4. Numerical experiments

The implementation of the kinematic formulation presented in Sections 2 and 3 has been carried out in MoDELib (Po et al., 2014). To illustrate the accuracy and efficacy of the implementation, a series of tests have been presented in this section. All the tests have been carried out using the material properties of single crystal copper. The methodology is also applicable for other crystal structures and complex crystal RVE volumes provided that the RVE crystal volume is a supercell of the underlying unit cell.

**4.1. Prediction of mean free path of dislocations under uniform expansion**

With the framework presented in Sec. 3.1, it is possible to predict the mean free path of the dislocation loops before self-annihilation under pure expansion. The theory yields the shift vectors, **Bξ**, whereas the value of the loop diameter at self-annihilation can be determined from the implementation. Figure 6 represents a plot of theoretical calculations for the self-annihilation distance, $\|\mathbf{B}\xi\|$, on the horizontal axis calculated using Eq. (39) and measured value of diameter of the periodic loop just before self-annihilation on the vertical axis. A linear plot establishes that the theoretical framework presented in this paper can determine the mean free path of dislocations under pure expansion for optimizing the size of the crystal RVEs. The mapped dislocation network segments in the crystal RVEs are also shown in Figure 6 for the visual confirmation of the dislocation density in the RVE prior to the first occurrence of self-annihilations due to periodic reentry. It was observed that a slight change in the aspect ratio of the simulation domain results in a drastic increase in the total number of periodically equivalent glide planes and hence results in a drastic increase in the mean free path of dislocations before self-annihilation.

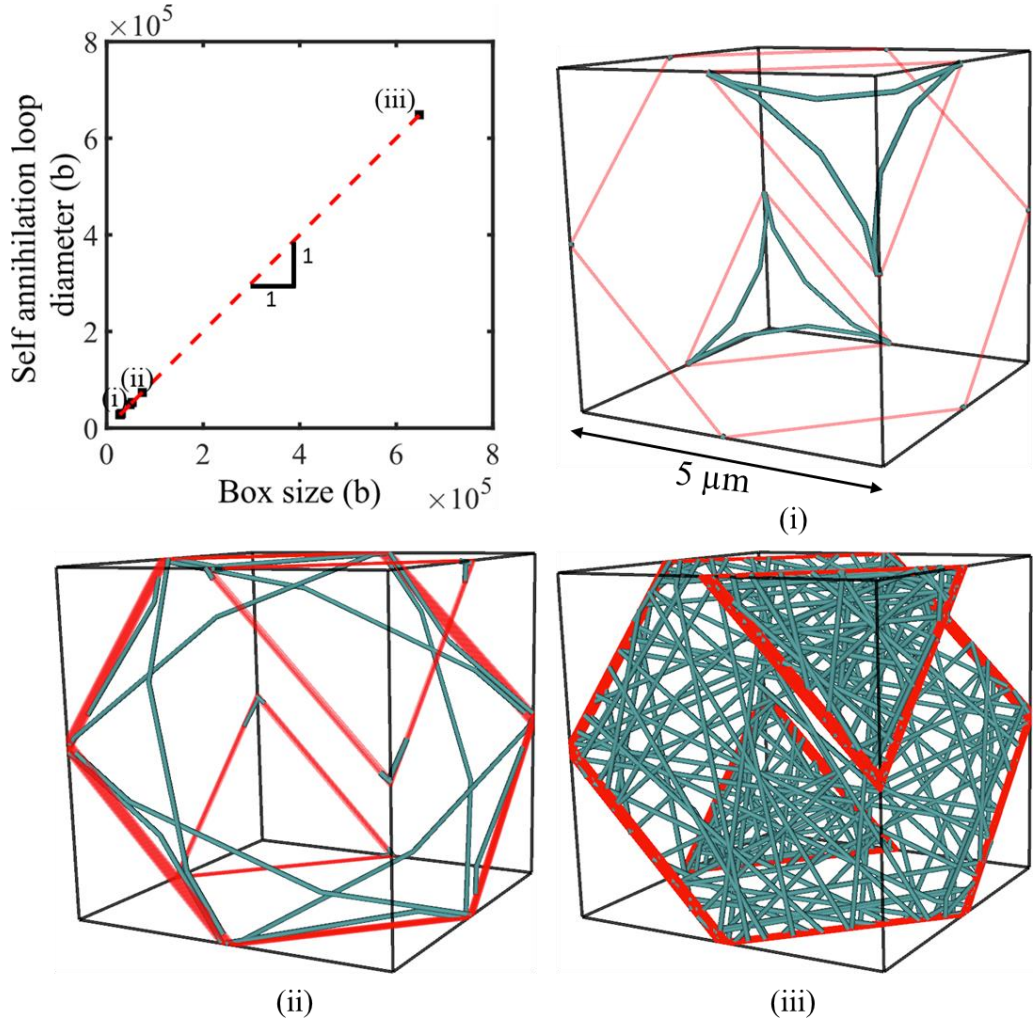

Figure 6: Theoretical shifts, $\|\mathbf{B}\boldsymbol{\xi}\|$, vs measured self-annihilation distance for a periodic loop in crystal RVE undergoing a uniform expansion, Aspect ratios of the RVE boxes are: (i) [1:1:1], (ii) [1:1.01:1.015], and (iii) [1:1.001:1.002]. Red lines in (i)-(iii) shows the glide planes of dislocation network segments. Thicker red lines show clusters of glide planes in proximity. The angle of the view does not allow to see that different parallel glide planes are created for skew aspect ratios.

## 4.2. RVE size invariance of loop kinematics

In this subsection, we demonstrate the kinematics of a dislocation loop under pure translation and uniform expansion. The accuracy of the kinematic implementation is tested for the two cases by changing the size of the RVEs as the kinematic behavior must not be affected by the size of the RVE. In other words, the kinematic implementation must be invariant of the presence of RVE boundaries. The size of the loop is kept constant for the two RVE sizes. These tests also address any instabilities in the numerical implementation when the dislocation segments are mapped back into the RVE.

Figure 7 presents the kinematics of a dislocation loop under pure translation for two different RVE sizes. (A pure translation in the glide plane is not a likely scenario of physical dislocation movement. However, it is used here for numerical demonstration purposes). As can be seen in the figure, the volumetric plastic distortion norm in the crystal is independent of the RVE size albeit the network configuration is different for the two RVEs. Depending on the position of the network segments, the glide planes are inserted dynamically and shifted to update bulk glide planes for the dislocation loop. Figure 7 (i)-(ii) represents the snapshots for the dislocation loop and network for an RVE size of 5 μm whereas Figure 7(iii)-(iv) represents the snapshots for the dislocation loop and network for an RVE size of 10 μm. No fluctuations in the plot of evolution of volumetric plastic distortion norm with time indicates that the implemented methodology is stable in the presence of RVE boundaries yet independent of the boundaries. Kinematics of the dislocation loop under pure expansion is similar to the kinematics of the loop under translation. The only difference is in the evolution of the norm of the volumetric plastic distortion (plastic distortion × volume), as shown in Figure 8, which is quadratic for a dislocation loop under pure expansion as opposed to constant for the loop under pure translation.

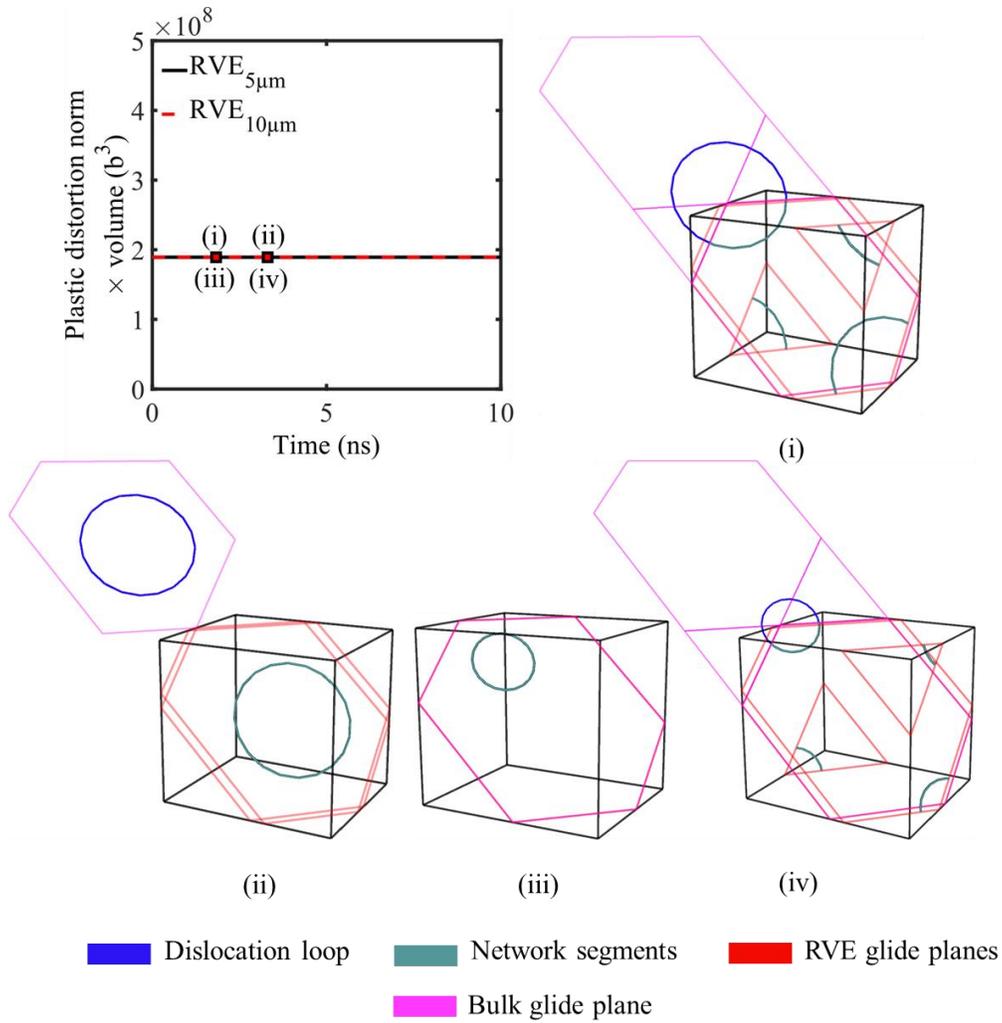

Figure 7: Evolution of volumetric plastic distortion norm with time for translation of a single dislocation loop in crystal RVE (The aspect ratio of the simulation box is [1:1.1:1.2]). Roman number snapshots indicate the evolution of the loop ((i) and (ii) for the RVE of 5 μm size and (iii) and (iv) for the RVE of 10 μm size and are marked in the plot of the volumetric plastic distortion norm).

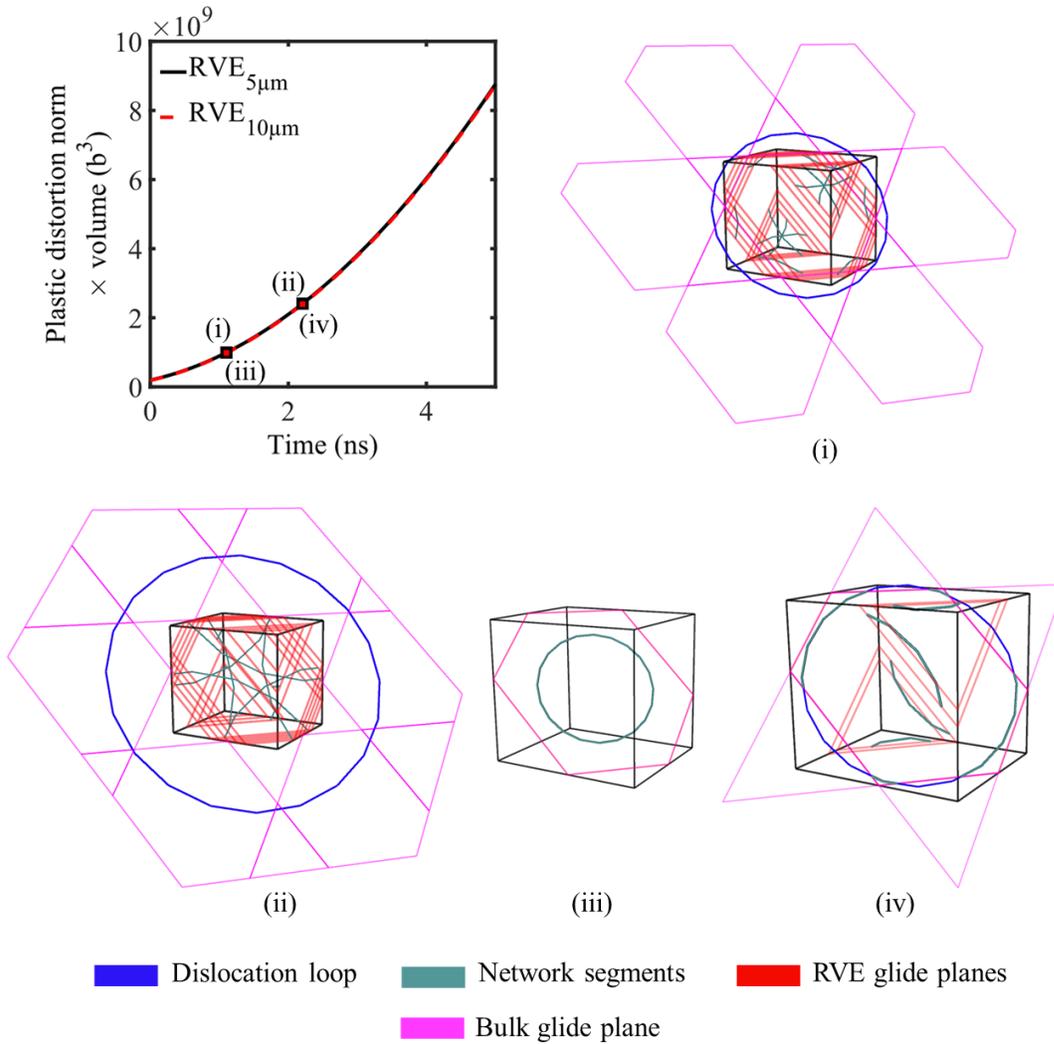

Figure 8: Evolution of volumetric plastic distortion norm with time for uniform expansion of a single dislocation loop in the crystal RVE (The aspect ratio of the simulation box is [1:1.1:1.2]). Roman number snapshots indicate the evolution of the two loops ((i) and (ii) for the RVE of 5 μm size and (iii) and (iv) for the RVE of 10 μm size and are marked in the plot of the volumetric plastic distortion norm).

### 4.3. Dislocation junction kinematics

During a junction formation, two dislocation network segments in the RVE combine to form a new network segment. The new network segment may belong to a different slip system as compared to the two reacting network segments. If the new network segment is glissile, a new loop of zero area is inserted to allow for the evolution of the dislocation density. This leaves behind a virtual density consisting of zero Burgers vector at the reacting site. The virtual density changes as the dislocation densities evolve. The dual topological framework allows for creation of glissile junctions across the boundary where the loops are created in the bulk crystal space and subsequently mapped into the RVE through the network segments. Cross slip is also executed in a similar fashion wherein a new loop is created if a screw segment is capable of gliding in the cross-slip plane. In this subsection, kinematics of two dislocation loops forming a sessile junction under pure translation and expansion is presented whereas the kinematic of a glissile junction will be presented in the next subsection.

Like the kinematics of single loop under translation, Figure 9 presents the translation kinematics of two dislocation loops forming a sessile junction. The junction length remains constant for the two loops under pure translation. On the other hand, Figure 10 presents the expansion kinematics of dislocation loops forming a sessile junction. The junction length increases linearly for the two loops under pure expansion. The presence of RVE boundaries do not affect the junction length during the reintroduction of dislocation segments in the RVE indicating the stability of junction kinematics in the RVE.

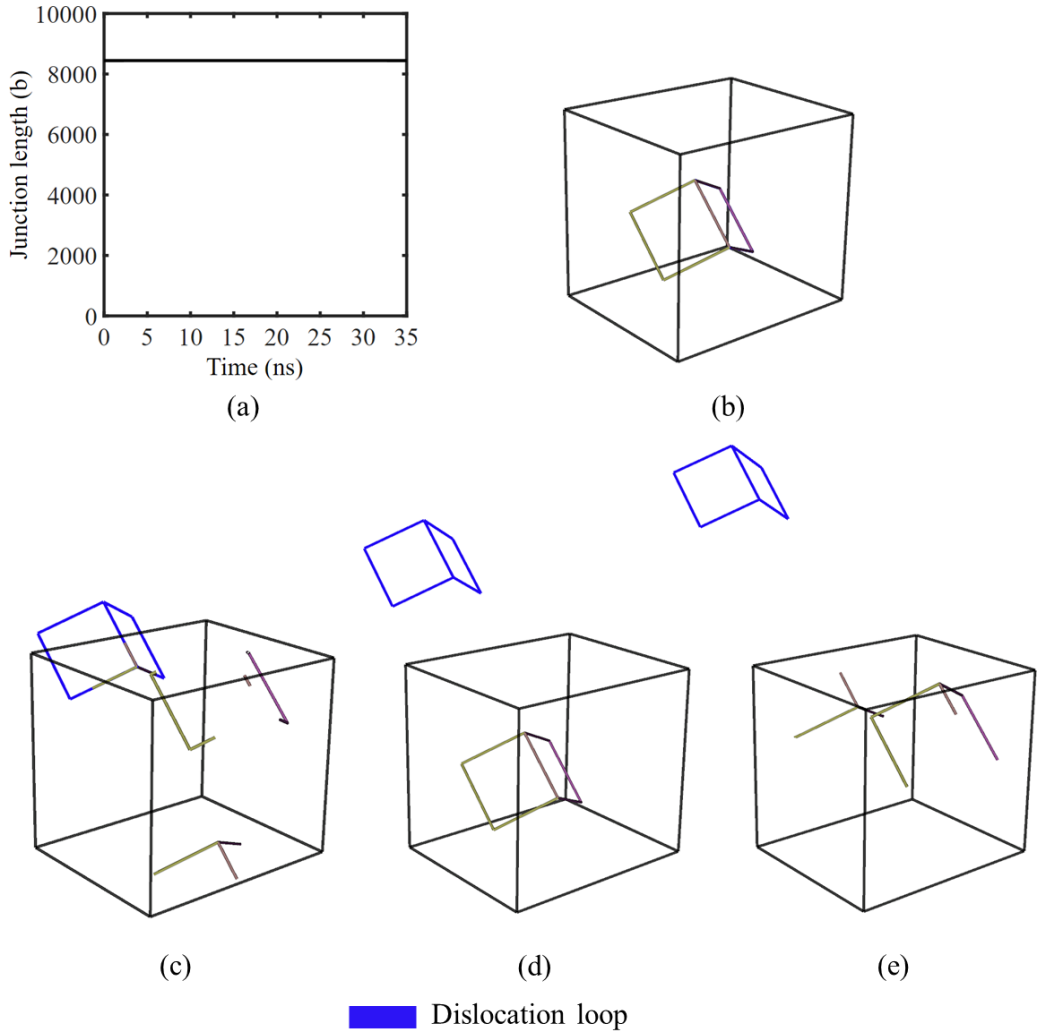

Figure 9: Translational kinematics of two loops forming a sessile junction in crystal RVE: (a) Evolution of junction length with time, and (b), (c), (d) and (e) shows the snapshots of translation of two loops forming the junction at different times. Dislocation segments within the RVE are the network segments. (The size of the RVE is approximately 5 μm. The aspect ratio of the simulation box is [1:1.1:1.2]).

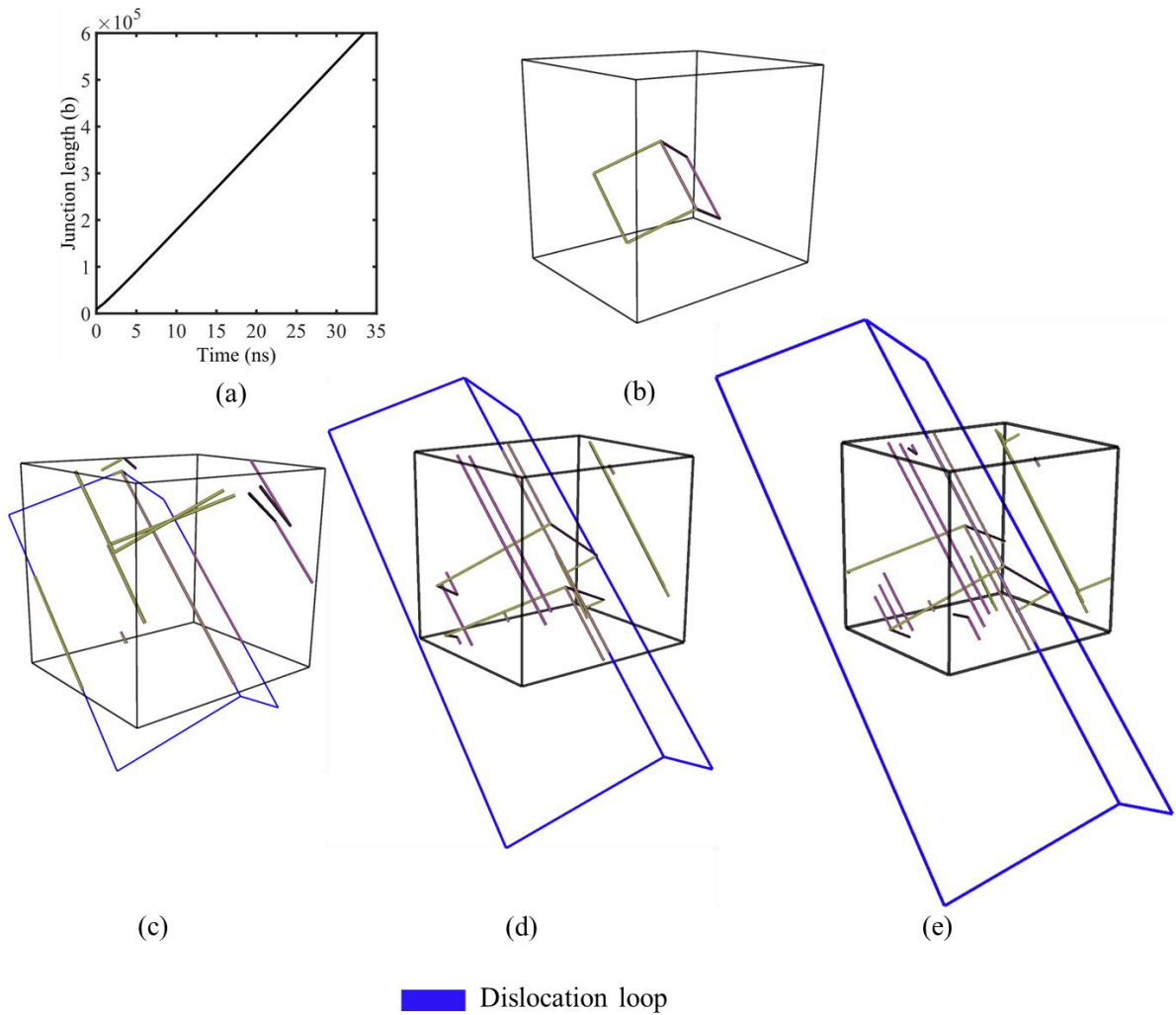

Figure 10: Uniform expansion kinematics of two loops forming a sessile junction in crystal RVE: (a) Evolution of junction length with time, and (b), (c), (d) and (e) shows the snapshots of translation of two loops forming the junction at different times. Dislocation segments within the RVE are the network segments. (The size of the RVE is approximately 5 μm. The aspect ratio of the simulation box is [1:1.1:1.2])

## 4.4. Translation-invariant network kinematics

Figure 11 presents the kinematics of a glissile junction using the dual topology in MoDELib. The evolution of a new glissile loop in a new slip system formed from reacting dislocation network segments in the RVE is shown in Figure 11(a)-(b) with Figure 11(a) representing the initial configuration and Figure 11(b) representing the evolution of dislocation densities after the formation of the glissile junction. To test the accuracy of kinematical framework for glissile junction formation, the junction configuration presented in Figure 11(a) is translated in their glide planes to start from different initial positions of the loops in the RVE. The kinematic evolution of the dislocation loops must be invariant of such translations. Figure 11(c) represents a translated variant of Figure 11(a) with Figure 11 (d) representing the evolution of the dislocation density in the translated variant after the formation of glissile junction. It is to be noted that the slip systems as well as the size of the loops remains the same for the translated variant. Figure 11(e)-(f) presents the results for the tests carried out for testing the accuracy of the implemented framework. The test was carried out for two different translated variants of the dislocation microstructure presented in Figure 11(a) out of which only one variant is shown in Figure 11(c). Figure 11(e) represents the evolution of norm of plastic distortion with time for the three cases whereas Figure 11(f) indicates the evolution of the slip system densities along with the relative errors for the two cases with respect to a reference configuration. The reference configuration is taken to be the dislocation configuration presented in Figure 11(a). Maximum relative errors of 0.35% and 0.08% were observed for case 1 and case 2 respectively with respect to the reference configuration, hence increasing the confidence in the implemented framework.

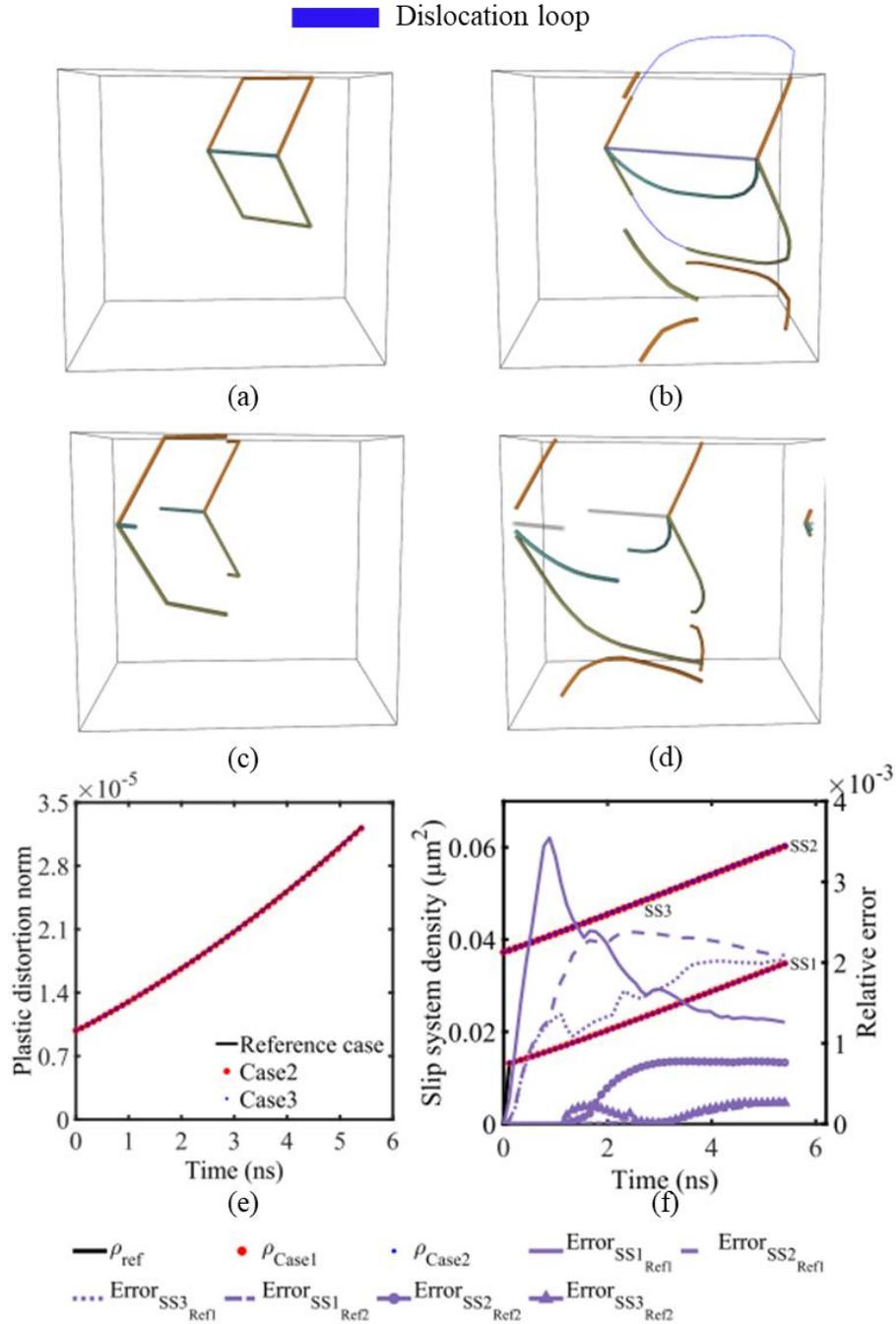

Figure 11: Kinematics of glissile junction in crystal RVE (Reference configuration is the dislocation microstructure represented by (a), Case 1 is the initial microstructure represented by (c), Case 2 is another translated version of the reference case (not shown in the figure)): (a) Initial dislocation microstructure for the reference configuration, (b) Final dislocation microstructure for the reference configuration, (c) Initial dislocation microstructure for the first translated configuration (Case 1), (d) Final dislocation microstructure for the first translated configuration (Case 1), (e) Evolution of norm of plastic distortion with time for the three cases, and (f) Evolution of slip system densities along with relative errors in the evolution of slip system densities with respect to the reference case (Label for slip systems: SS1-$(\bar{1}1\bar{1})/[0\bar{1}\bar{1}]$, SS2-$(\bar{1}1\bar{1})/[110]$, and SS3- $(\bar{1}\bar{1}\bar{1})/[10\bar{1}])$. Dislocation segments within the RVE are the network segments.

### 4.5. Bulk network kinematics

The implementation of the dual topological representation of dislocations is also tested for a typical DDD calculation incorporating discrete glide events such as junction formation, cross slip and adaptive remeshing of the dislocation networks. Circular loops of radius 4 µm were populated randomly in 12 slip-systems of a Cu crystal of size 5 µm × 5.5 µm × 5 µm such that the initial density is $3 \times 10^{12}$ m$^{-2}$ and the crystal was loaded at a strain rate of $100$ s$^{-1}$. Other details for the bulk simulation have been omitted as a detailed description of the dynamics of the dislocation loop will be a subject of a follow-up publication. Figure 12 represents the evolution of the dislocation density with Figure 12(a) representing the initial dislocation density of approximately $3 \times 10^{12}$ m$^{-2}$ and Figure 12(b) of approximately $1.2 \times 10^{13}$ m$^{-2}$, which is a 4 times increase in the dislocation density. Figure 12(c) presents the evolution of glissile, sessile and the total dislocation density up to 0.245% strain. During deformation of the crystal, a loop can form sessile junctions or glissile junctions and cross-slip to a conjugate plane leaving behind virtual densities. With the dual topological framework presented in this paper, it becomes much easier to track the plastic distortion as well as acquire the statistics of the formation of junctions as well as cross slip from static dislocation configurations. It is believed that the presented methodology is a one-step forward towards acquiring the statistical based description of the dislocation from DDD to be utilized for larger scale simulations.

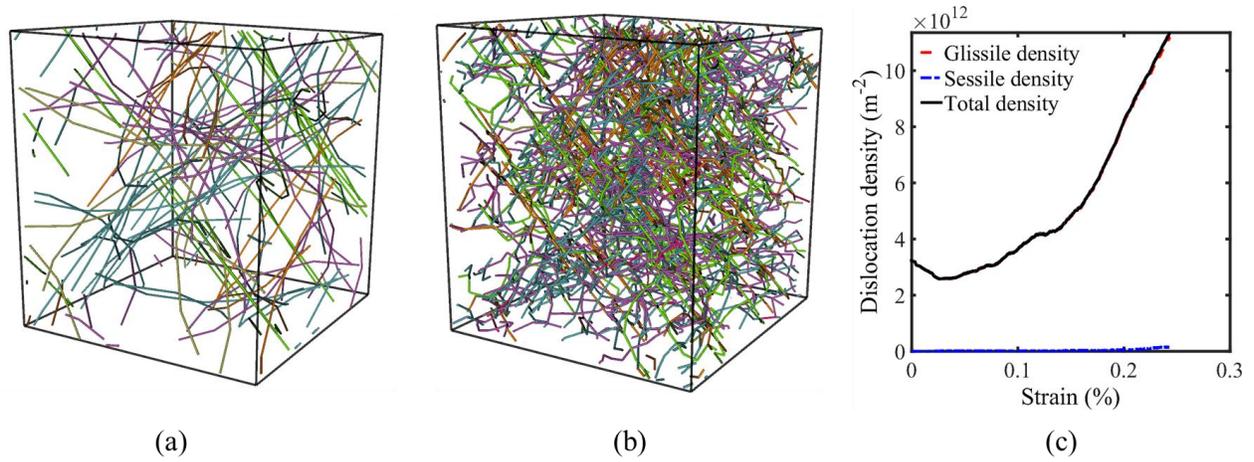

(a)                  (b)                  (c)

Figure 12: Evolution of dislocation density during bulk simulations: (a) Initial dislocation density for the bulk crystal simulation, (b) Dislocation density for the bulk crystal simulation at a strain of 0.24%, and (d) Evolution of glissile, sessile and total dislocation density.

## 5. Summary and conclusions

Statistical homogeneity-based treatment of dislocation flux boundary condition for discrete dislocation loops and networks in a representative volume element has been addressed from the standpoint of bulk crystal plasticity modeling. An effective and robust approach to model mesoscale plasticity has been presented using a dual topological framework. An object-based representation of the glide planes in the crystal RVE and the presence of commensurate lattice allows for setting up a mapping between the RVE crystal space and the bulk crystal space. Dislocation loops move in bulk glide planes in the bulk crystal space whereas dislocation network segments move in the glide planes in the RVE space. Bulk glide planes can be constructed from the RVE glide planes through periodic shifts in the RVE. The plastic distortion associated with the dislocation loops is available at all time steps through a surjective mapping between the bulk crystal space and the RVE space. Subsequently, testing of the framework for discrete glide events such as junction formation as well as bulk crystal loading established the accuracy of the implementation for the kinematic realization of the flux boundary conditions.

With the current framework, boundary nodes for dislocation segments do not contribute towards the motion of those segments. In other words, within a periodic framework, the boundary nodes do not bear any physical relevance for solving the equation of motion of the dislocation network. The framework presented in this paper is expected to be pivotal for the development of continuum-based theory of dislocations.

## Appendix

A space-filling volume can be obtained as the Voronoi cell of a 3D lattice. Let the primitive vectors of this lattice be $\mathbf{b}_1, \mathbf{b}_2, \mathbf{b}_3$. Consider two parallel planes in this 3D lattice, the first plane with normal $\hat{\mathbf{m}}$ and passing through $\mathbf{x}_1$, and the second plane passing through $\mathbf{x}_2$. The two planes are periodically equivalent if plane 2 contains a periodic image of an arbitrary point $\mathbf{x}$ on plane 1. A periodic image of $\mathbf{x}$ has position

$$\mathbf{x}' = \mathbf{x} + \xi_1 \mathbf{b}_1 + \xi_2 \mathbf{b}_2 + \xi_3 \mathbf{b}_3 = \mathbf{x} + \mathbf{B}\boldsymbol{\xi}, \quad (A.1)$$

where $\mathbf{B}$ is the matrix with columns corresponding to the vectors $\mathbf{b}_i$, and $\boldsymbol{\xi}$ is a column vector of integers. The requirement of periodic equivalence of the two planes then results in the conditions

$$\begin{cases} \hat{\mathbf{m}} \cdot (\mathbf{x} - \mathbf{x}_1) = 0 \\ \hat{\mathbf{m}} \cdot (\mathbf{x} + \mathbf{B}\boldsymbol{\xi} - \mathbf{x}_2) = 0 \end{cases}. \quad (A.2)$$

Eliminating $\hat{\mathbf{m}} \cdot \mathbf{x}$ yields the condition of periodic equivalence of the two planes as.

$$\hat{\mathbf{m}} \cdot (\mathbf{x}_1 - \mathbf{x}_2 + \mathbf{B}\boldsymbol{\xi}) = 0 \quad \text{for some integer } \boldsymbol{\xi}. \quad (A.3)$$

In general, Eq. (A. 3) cannot be satisfied for a vector of integers $\boldsymbol{\xi}$. This is because no special relationship exists between the plane normal $\hat{\mathbf{m}}$ and the vectors $\mathbf{b}_i$. However, the general case bears little relevance in the analysis of crystal RVEs. The case relevant to crystal RVEs is that where $\hat{\mathbf{m}}$ is the normal to lattice planes of the underlying crystal lattice, and where the lattice vectors $\mathbf{b}_i$ span a sub-lattice of the crystal lattice. Letting $\mathbf{a}_1, \mathbf{a}_2, \mathbf{a}_3$ be a basis of the crystal lattice, and $\mathbf{A}$ be the matrix having these vectors in columns, the case under consideration can be expressed by the following conditions:

$$\mathbf{B} = \mathbf{AN} \tag{A. 4}$$

$$\mathbf{x}_1 = \mathbf{A}\boldsymbol{\xi}_1 \tag{A. 5}$$

$$\mathbf{x}_2 = \mathbf{A}\boldsymbol{\xi}_2 \tag{A. 6}$$

$$\hat{\mathbf{m}} = \frac{\mathbf{A}^{-T}\boldsymbol{\zeta}}{\|\mathbf{A}^{-T}\boldsymbol{\zeta}\|}, \tag{A. 7}$$

where $\boldsymbol{\xi}_1, \boldsymbol{\xi}_2$ and $\boldsymbol{\zeta}$ are integers while $\mathbf{N}$ is an integer matrix. In this case, the condition of plane periodic equivalence (A. 3) becomes

$$\boldsymbol{\zeta}^T \mathbf{A}^{-1}(\mathbf{A}\boldsymbol{\xi}_1 - \mathbf{A}\boldsymbol{\xi}_2 + \mathbf{AN}\boldsymbol{\xi}) = 0 \qquad \text{for some integer } \boldsymbol{\xi} \tag{A. 8}$$

or

$$\boldsymbol{\zeta}^T(\mathbf{N}\boldsymbol{\xi}) = \boldsymbol{\zeta}^T(\boldsymbol{\xi}_1 - \boldsymbol{\xi}_2) \qquad \text{for some integer } \boldsymbol{\xi}. \tag{A. 9}$$

By virtue of Bézout's identity, this equation admits integer solutions if and only if the integer $\boldsymbol{\zeta}^T(\boldsymbol{\xi}_1 - \boldsymbol{\xi}_2)$ is a multiple of the greatest common divisor (GCD) of elements of the integer vector $\boldsymbol{\zeta}^T\mathbf{N}$. The quantity

$$n_p(\boldsymbol{\xi}, \mathbf{N}) = \text{GCD}(\boldsymbol{\zeta}^T\mathbf{N}) \tag{A. 10}$$

is in fact the smallest integer that can be written in the form $\boldsymbol{\zeta}^T(\boldsymbol{\xi}_1 - \boldsymbol{\xi}_2)$, and any other integer of this form is a multiple of $n_p(\boldsymbol{\xi}, \mathbf{N})$. The integer $n_p(\boldsymbol{\xi}, \mathbf{N})$ is therefore the number of lattice planes that separates periodically equivalent planes with normal $\hat{\mathbf{m}} = \frac{\mathbf{A}^{-T}\boldsymbol{\zeta}}{\|\mathbf{A}^{-T}\boldsymbol{\zeta}\|}$ in a domain defined by the Voronoi cell of the sub-lattice with basis defined by the columns of $\mathbf{B} = \mathbf{AN}$. Because the interplanar distance of that family of planes is $d_{min}(\boldsymbol{\zeta}) = \frac{1}{\|\mathbf{A}^{-T}\boldsymbol{\zeta}\|}$, then the distance between periodically equivalent planes is

$$d_p(\boldsymbol{\zeta}, \mathbf{N}) = d_{min} n_p = \frac{1}{\|\mathbf{A}^{-T}\boldsymbol{\zeta}\|} GCD(\boldsymbol{\zeta}^T \mathbf{N}). \tag{A. 11}$$

We are interested in finding the total number of periodically equivalent lattice planes in a crystal RVE. The projection of the sides of the box along the plane normal can be represented by

$$\mathbf{P}_m = \frac{\mathbf{m}^T \mathbf{B}}{|\mathbf{m}|}. \tag{A. 12}$$

Substituting $\mathbf{m} = \mathbf{A}^{-T}\boldsymbol{\zeta}$ and $\mathbf{B} = \mathbf{AN}$, $\mathbf{P}_m$ can be represented as

$$\mathbf{P}_m = \frac{\boldsymbol{\zeta}^T \mathbf{N}}{|\mathbf{m}|}. \tag{A. 13}$$

In order for the projection of the RVE box to be representing periodically equivalent lattice planes, $\mathbf{P}_m$ should be a multiple of $GCD(\boldsymbol{\zeta}^T \mathbf{N})$, i.e. $\frac{\boldsymbol{\zeta}^T \mathbf{N}}{|\mathbf{m}|} = \mathbf{K}\frac{GCD(\boldsymbol{\zeta}^T \mathbf{N})}{|\mathbf{m}|}$. This implies that

$$\mathbf{K} = \frac{\boldsymbol{\zeta}^T \mathbf{N}}{GCD(\boldsymbol{\zeta}^T \mathbf{N})}, \tag{A. 14}$$

where $\mathbf{K}$ is a $3 \times 1$ vector representing the number of periodically equivalent planes along each direction of the box. Hence, the total number of periodicaly equivalent planes can be given by

$$N_p^{eq} = \text{sum}(\mathbf{K}), \tag{A. 15}$$

where the sum represents the sum of each individual element in the vector $\mathbf{K}$. Equation (A. 15) gives the information about the total number of periodic equivalent planes in a crystal RVE. But it does not give a measure of the self-annihilation distances of the periodic loops. To determine the measure for the self-annihilation distance of a periodic loop, consider two periodically equivalent lattice planes in a crystal RVE as defined by (A. 3). Self-annihilation of a periodic loop can be represented physically by the two lattice points becoming same in the crystal RVE. Mathematically, this can be represented as

$$\boldsymbol{\zeta}^{\mathrm{T}}(\mathbf{N}\boldsymbol{\xi}) = 0. \tag{A. 16}$$

Eq. (A. 16) represents a Diophantine equation with integer solutions. Hence, for the case of self-annihilation of the periodic loop, $\boldsymbol{\xi}$ gives the integer shifts when the periodic loops will cross each other resulting in self-annihilation. It is worthwhile to note that one of the general solutions of the Eq. (A. 16) is the shift vector $(0,0,0)$ which represents the starting configuration. Hence, the closest solution to $(0,0,0)$ for the diophantine equation will give a measure of the self-annihilation of the periodic loop.

**Acknowledgements**

A. El-Azab acknowledges support from the U. S. Department of Energy, Division of Materials Sciences and Engineering, through award number DE-SC0017718 at Purdue University. Y. Pachaury was supported by the US Department of Energy, Office of Nuclear Energy, contract DE-NE0008758 at Purdue University. G. Po acknowledges the support of the U.S. Department of Energy, Office of Fusion Energy Sciences (FES), under award number DE-SC0019157 with UCLA, and subaward number 019GXA906 with the University of Miami.